\documentstyle[multicol,aps,prb,epsf,graphics,psfig]{revtex}
\input epsf

\newcommand{\be}{\begin{equation}}
\newcommand{\bel}[1]{\begin{equation}\label{#1}}
\newcommand{\ee}{\end{equation}}
\newcommand{\bea}{\begin{eqnarray}}
\newcommand{\ba}{\begin{array}}
\newcommand{\eea}{\end{eqnarray}}
\newcommand{\ea}{\end{array}}

\begin{document}
\draft

\title{Boundary-bulk interplay of molecular motor traffic flow through a compartment}
\author{M. Ebrahim Fouladvand (1,2), Modjtaba Salehi (1) and Mostafa Yadegari (1)}

\address{(1) Department of Physics, Zanjan University, P.O. Box
45196-313, Zanjan, Iran.\\
(2) Department of Nano-Science, Institute for Studies in
Theoretical Physics and Mathematics (IPM),\\
 P.O. Box 19395-5531, Tehran, Iran. }

\maketitle

\begin{abstract}

The flow of motor proteins on a filamental track is modelled
within the the framework of lattice driven diffusive systems.
Motors, considered as hopping particles, perform a highly biased
asymmetric exclusion process when bound to the filament. With
a certain rate, they detach from the filament and execute unbiased
random walk in the bulk which is considered as a closed cubic
compartment. Motors are injected (extracted) from the leftmost
(rightmost) site of the filament located along the symmetry axis
of the compartment. We explore the transport properties of this
system and investigate the bulk-boundary interplay on the system
stationary states. It is shown that the detachment rate  notably
affects the system properties. In particular and in contrast to
ASEP, it is shown that the density profile of bound particles
exhibit different types of non monotonic behaviours when the
detachment rate varies. It is shown that in certain situations,
the density profile of the filament consists of coexisting high
and low regions.

\end{abstract}
\pacs{PACS numbers: 02.50.Ey, 05.40.-a, 05.70.Ln, 64.60.-i } ]
\begin{multicols}{2}

\section{Introduction}

Many transport problems in physics and related interdisciplinary
areas have been successfully modelled within the framework of
lattice driven diffusive systems. The kinetics of
biopolymerization \cite{ligget,zia,schutz1}, interface
growth,\cite{zia,hinrischsen} and vehicular traffic flow
\cite{css99,helbing,kernerbook} are among various examples of
diverse applications of cellular automaton approach in describing
out of equilibrium systems. Driven lattice gases with open
boundaries are stochastic lattice models of particles which hop
preferentially in one direction and which are connected at their
boundaries to particle reservoirs of constant densities. As a
result of this coupling, a stationary flow of particles is
established in the system. A prototype example of non-equilibrium
low dimensional systems is the asymmetric simple exclusion process
(ASEP) where particles interact through hard-core repulsion
\cite{mcdonald}. ASEP has been applied to describe a variety of
phenomena in physics and beyond. In one dimension, many
interesting feature such as boundary-induced phase transitions,
spontaneous symmetry breaking and shock formation have been shown
to exist \cite{krug,derida,domany,godrech,schutz2}. These
phenomena are associated to the boundary effects on the chain bulk
and are generally termed {\it boundary induced}. Besides the
boundaries, the bulk properties of the 1D chain can be affected
via coupling to a $3D$ bulk reservoir. A well known example is the
{\it Langmuir kinetics } (LK) \cite{langmuir} where particles can
adsorb (desorb) if a bulk site is empty (occupied). Recently the
interplay of both effects i.e., the boundary and the bulk
reservoirs have received attention
\cite{frey1,frey2,evans1,juhasz}. It has been verified that in
certain situations where the adsorption/desopbtion rate properly
scales with the system size, the density profile exhibits a
non-monotonic structure in the bulk which is characterized by the
coexistence of low and high density regions separated by a shock
front if the system size is high enough. In contrast to ASEP where
shocks are not localized, in the presence of adsorption/desorption
of particles, the system shows a more complex and richer
behaviour. Specially, the shocks are localized. This behaviour can
be verified by a continuous mean-field approach
\cite{frey1,frey2,evans1}. Along this line of study, we
investigate the bulk-boundary interplay in a one dimensional
chain. In our study, the homogeneous bulk reservoir is replaced by
a more realistic structure. More specifically, the bound particles
can detach from the one dimensional chain and execute unbiased
random walk. The unbound particles are confined to a compartment
with reflecting boundaries. It is one of our main objectives to
see how the cooperative behaviours of the diffusive unbound
particles affects the results obtained from the static homogeneous
3D bulk. We are interested to know if the shock localisation persists
in the case where the bulk particles have dynamics. Our study is
motivated from a biological application of ASEP. A considerable
portion of intra-cell transport of biological cargos in eucaryotic
cells is performed via the so-called motor proteins namely
kinesin, myosin and dynein which move on filamentary tracks called
microtubules and actins
\cite{howard,schliwa,ajdari1,ajdari2,nedelec1,schnitzer,okada,nedelec2}.
These bio motors perform a biased processive random walk on the
filament and after some while detach from it due to thermal noise.
Recently some physicists have attempted to formulated the problem
described above within the lattice driven gas framework
\cite{klump1,klump2,klump3,klump4,klump5}. Besides investigating
the diffusion properties of a single molecular motor on a
filament-like track, these investigations have been generalized to
account for the mutual interaction of motors via hard core
repulsion. The interaction was considered at its simplest form
i.e., {\it exclusion}. It has been shown that anisotropy induced
by the filament can affect the diffusion properties of the bulk
particles which are unbound to the filament. On the other hand,
the bulk diffusive motors can substantially influence the
structure of the density and current profiles on the filament. It
has been recently been shown \cite{klump6} that in a compartment
containing  cytoskeletal filaments, hard core interaction between
unbound motors, which are confined to move inside the compartment,
and the filament gives rise to a localized shock structure in the
density profile of the filament. The above geometrical arrangement
i.e., a filament placed along the symmetry axis of a compartment
are accessible to {\it in vitro} experiments and mimics the
structure observed in cytoskleton. In particular, such systems
serve as simple models for the description of motor-based
transport in an axon \cite{goldstein,dahlstrom}. In this paper our
main focus is on the role of detachment rate on the filament
density profile. The investigation of the consequences of changing
the detachment rate can be important not only from biological but
also for theoretical purposes.

\section{ Description of the Problem and its formulation}

Before proceeding further into the problem, let us define the
model in more details. The motor proteins motion is assumed to
take place in a compartment. For simplicity, we take the
compartment shape a three dimensional cube with dimensions $Lx,
Ly$ and $Lz$ in the directions $x,y$ and $z$ respectively. We
assume the filament is located along the $x$ axis of the tube.
Furthermore, the filament coordinates in $y$ and $z$ directions
are taken as $y=\frac{Ly}{2}$ and $z=\frac{Lz}{2}$. We assume the
particles (motors) enter onto the filament from the left boundary
i.e., at the point $(x,y,z)=(0,\frac{Ly}{2},\frac{Lz}{2})$. The
particles can leave the compartment only via reaching the last
point of the filament which is
$(x,y,z)=(Lx,\frac{Ly}{2},\frac{Lz}{2})$. The particles movement
are governed according the following discrete-time stochastic
dynamical rules. At each time step $t$, the state of the system is
specified by the occupation numbers of each site inside the
compartment. If the site is full, its occupation number is one. If
it is empty, the occupation number is zero. In addition, the
motors are assumed to be point particles.  Note that the integer
coordinates $i,j$ and $k$ of the sites are restricted as $0 \leq i
\leq Lx$;~~$0 \leq j \leq Ly$;~~$0 \leq k \leq Lz$. Now we explain
the updating rules applied to all the particles inside the compartment.\\

1) {\bf particle injection}:\\

In the beginning of the time step, we update the entry site of the
filament: if the first site of the filament is empty, a particle
will be injected to it with the probability $\alpha$. Furthermore,
The number of entrants will be regarded as the system inflow.\\

2) {\bf bound particle movement}:\\

We divide the particles inside the compartment into two groups:
bound and unbound. If the particle is on the filament, we consider
it as a bound one. If it is outside the filament, it is regard as
an unbound particle. As discussed, the filament role is to provide
a driven track on which the particles diffusive motion is highly
directed. A particle on the filament can randomly move one step
forward with probability $p$, one step backward with probability
$\delta$, remains immobile at its position with probability
$\gamma$ or leave the filament (detachment) to one of its
non-filament neighbours with probability $\frac{\epsilon}{6}$
\cite{klump3,klump4}. All these random movements will be
successful if the target site is empty. In case the target site is
already occupied by another motor, the attempted hop will be
rejected. The probabilities should sum up to one:
$p+\delta+\gamma+\frac{2\epsilon}{3}=1$. In three dimensions,
there are four neighbouring non-filament sites adjacent to each
filament site $(i,\frac{Ly}{2},\frac{Lz}{2})$ therefore the
detachment probability equals $\frac{2\epsilon}{3}$.\\

3) {\bf unbound particle movement}:\\

An unbound particle is the one which is not located on the
filament. These particles diffuse freely in the compartment space
until they find a chance to attach to the filament. Once they
attach to the filament, they become bound. The diffusion
probabilities to the adjacent empty sites are equally taken to be
$\frac{1}{6}$ for all six spatial directions. For unbound
particles, the compartment boundaries are assumed to be totally
reflective. This can be implemented by putting immobile particles
on the all boundary sites expect the entrance site
$(0,\frac{Ly}{2},\frac{Lz}{2})$ and the exit site $(Lx,\frac{Ly}{2},\frac{Lz}{2})$.\\

4) {\bf particle extraction}:\\

Once a bound particle reaches the last filament site, it will be
removed from the system (compartment) with the exit probability
$\beta$. The number of removed particles will be regarded as the
system outflow. We recall that due to binding/unbinding of
particles from the filament, in the steady state the current
depends on the filament site number.
Throughout this paper by current we mean the current of the last filament site.\\

{\bf updating scheme}:\\

 Before proceeding further, it would be illustrative to discuss,
in some details, the type of our updating. Basically, in the CA
(cellular automata) models, the updating schemes are divided into
two categories: {\it particle-oriented} and {\it site-oriented}.
The particle oriented scheme can itself be divided into various
methods. The most important one is parallel dynamics in which the
updating rules are synchronously applied to all particles. The
prototype example is the vehicular traffic flow \cite{css99}. The
next variant is the ordered sequential updating. In this scheme,
one updates the particles in an ordered sequential manner say from
left to right in one dimensional open systems. Finally the third
type is random sequential update. In this scheme, an update step
is composed of many sub-update steps. In each sub-update step, we
randomly chose one of the particles and update its position in
accordance to the movement rules. In contrast to particle-oriented
schemes, one has site-oriented ones. Site-oriented schemes, are
generically divided into ordered and random sequential types. In
the ordered scheme, one sweeps the lattice in a certain direction
(prescribed by the method) and updates the position of those sites
which are occupied by particles. In random sequential type, one
randomly selects a site. If this site is occupied, then updating
is done otherwise one continues by selecting other sites until an
occupied site is found. The type of updating will affect some
aspects of the problem but normally the phase structure of the
model will not substantially changed under changing the updating
scheme. In this paper we use particle-oriented ordered sequential
scheme. More specifically, we label each particle upon entering
into the compartment. At each time step, we update the particles'
positions in the order of their labels. This resembles the
site-ordered random sequential update. Our motivation to choose
this update stems from our interest to evaluate the passage time
of particles from entrance to  removal. In the following sections
we present our simulation results and discuss the phase structure
and the other statistical properties of this system.

\section{simulation results}

If we allow time to evolve for a long time we see the systems
reaches a non-equilibrium current-carrying steady state in which
the statistical properties of the system does not show significant
changes with time. To see this more explicitly, let us first
define the quantities we are interested to compute:\\

1) number of bound particles at the end of timestep $t$ denoted by
$N_b(t)$. The subscript $b$ denotes {\it bound}. Dividing $N_b(t)$
by the filament size $Lx$ gives the instantaneous global filament
density at time $t$ which we denote it by $\rho_b(t)$.\\

2) number of unbound particles at the end of timestep $t$ denoted
by $N_{ub}(t)$. Dividing $N_{ub}(t)$ by the compartment volume
$Lx \times Ly \times Lz-Lx$ gives the unbound global particle density at time $t$ which we denote it by $\rho_{ub}(t)$.\\

3) filament current $J_b(t)$ which is defined to be the number of
particle exited from the last filament site by the end of $t$-th
step divided by $t$.\\

4) passage time of each particle: since we label the particles
upon entering into the system, we can follow their trajectories
from the time they enter the compartment until the time they leave
it. Therefore we can specify each particle's passage time. $p(t)$
is defined to be the average passage time of those particles which
have already exited the compartment by the end of time $t$.\\

5) centre of mass of the unbound particles denoted by
$<X_{ub}>(t)$. This quantity is obtained by averaging over the
$x$-coordinate of all unbound particles at time $t$.\\

6) centre of mass of the bound particles denoted by $<X_b>(t)$.
This quantity is obtained by averaging over the
$x$-coordinate of all bound particles at time $t$.\\

7) total number of particles in the system i.e., $N_{ub}(t)+
N_b(t)$ which is named {\it load}.\\

We note that in the steady state, the radial current of particles
onto the filament should be equal to the current of particles
leaving the filament. This is called {\it radial equilibrium}
discussed in \cite{klump3,klump4}. In a mean-field approximation,
the radial equilibrium implies the following relationship between
$\rho_b$ and $\rho_{ub}$: \be
\frac{\epsilon}{6}\rho_b(1-\rho_{ub})=\frac{1}{6}\rho_{ub}(1-\rho_b)
\ee In the following we exhibit the stationary characteristics of
the system. The system contains $Lx=100$ sites. The perpendicular
dimensions are $Ly=Lz=25$ and $\delta=\gamma=0$ unless otherwise
specified. In the first figure, we sketch the dependence of
steady-state value of load in terms of $\alpha$ for three values
of detachment rate $\epsilon$ and exit rate $\beta$.

\begin{figure}\label{Fig1}
\epsfxsize=8truecm \centerline{\epsfbox{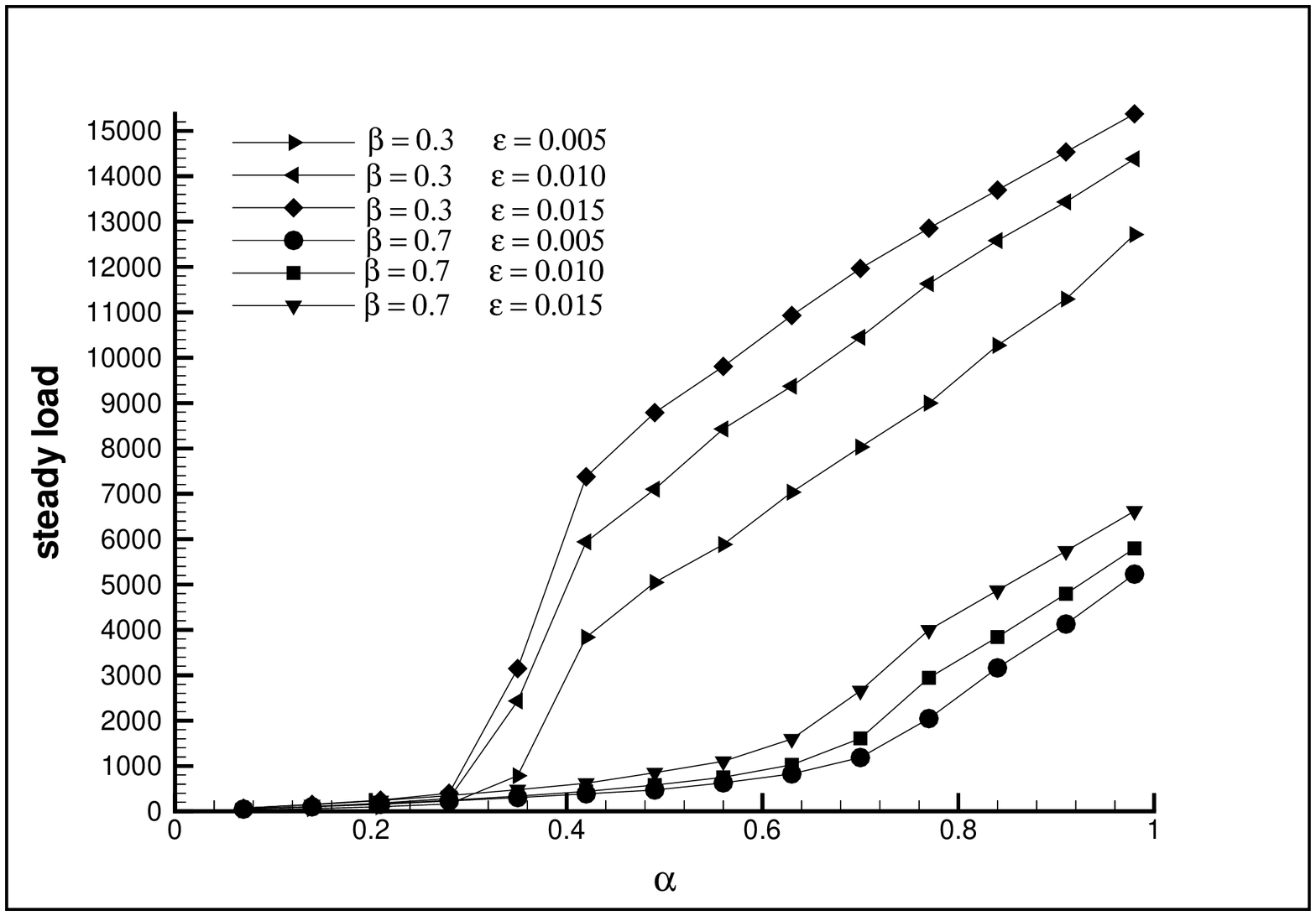}}
\end{figure}
\vspace{0.02 cm} {\small{Fig.~1: steady state value of load vs
$\alpha$. $\epsilon$ and $\beta$ are specified in the figure.} }\\

We observe a phase-transition like behaviour near $\alpha_c \sim
0.3$ and $0.7$, for $\beta=0.3$ and $0.7$ respectively, where the
steady load shows up a sharp increase. Moreover, increasing
$\epsilon$ gives rise to increment of the load. This is expected
since increasing $\epsilon$ raises the detachment probability
hence the number of unbound particles is increased. To verify this
transition, let us consider the dependence of the bound current
$J_b$ on $\alpha$.

\begin{figure}\label{Fig2}
\epsfxsize=8truecm \centerline{\epsfbox{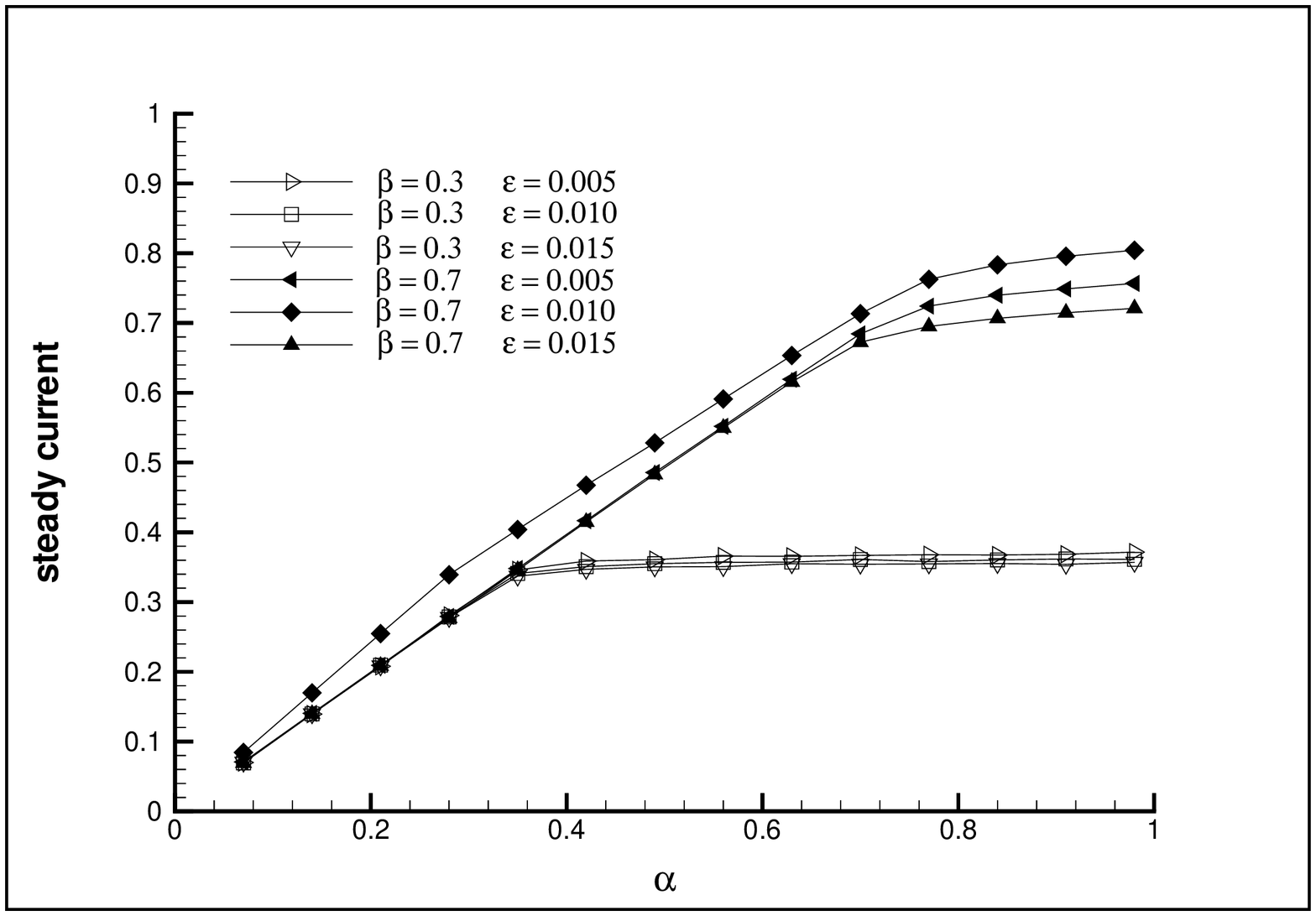}}
\end{figure}
\vspace{0.02 cm} {\small{Fig.~2: steady state current in terms of
$\alpha$ for a system of size $Lx=100$. The value of $\beta$ and
$\epsilon$ are specified in the figure.} }\\

for $\beta=0.3$ the current linearly increases with respect to
$\alpha$ and then saturates at an $\epsilon$-dependent value. The
saturation value lies near $0.35$.  The smaller values of
$\epsilon$ correspond to higher currents. This can be explained on
account of the fact that for small $\epsilon$ motors prefer to
remain bound to the filament track and continue their biased
movement toward the exit point. Furthermore, we observe that the
effect of increasing $\epsilon$ is smoothing the transition. For
larger value of $\beta=0.7$, the overall picture remains the same.
However the transition point and saturation value changes
correspondingly. These behaviour are analogous to ASEP. The next
quantity is the passage time. We exhibit the steady state passage
time vs $\alpha$ for some values of $\beta$ and $\epsilon$.

\begin{figure}\label{Fig3}
\epsfxsize=8truecm \centerline{\epsfbox{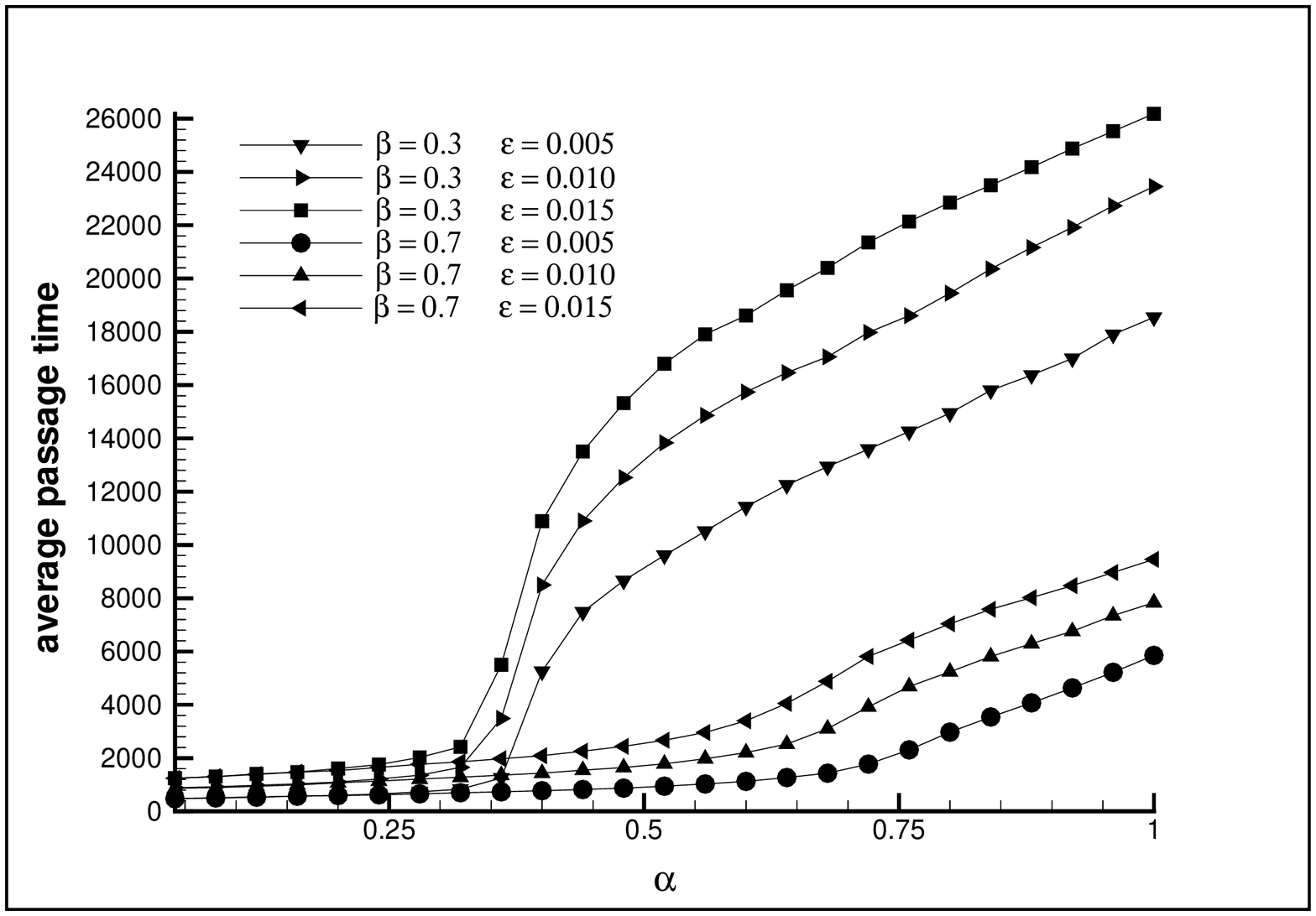}}
\end{figure}
\vspace{0.02 cm} {\small{Fig.~3: steady state passage time in
terms of $\alpha$. $Lx=100$ and the values of
$\beta$ and $\epsilon$ are specified in the figure.} }\\

The above graph is similar to the load graph. One can see the
phase transition is also manifested in the passage time. We note
that transition is smooth for $\beta=0.7$ and it resembles to a
cross over. Let us next consider the density of bound motors
$\rho_b$ as a function of $\alpha$.\\

\begin{figure}\label{Fig4}
\epsfxsize=8truecm \centerline{\epsfbox{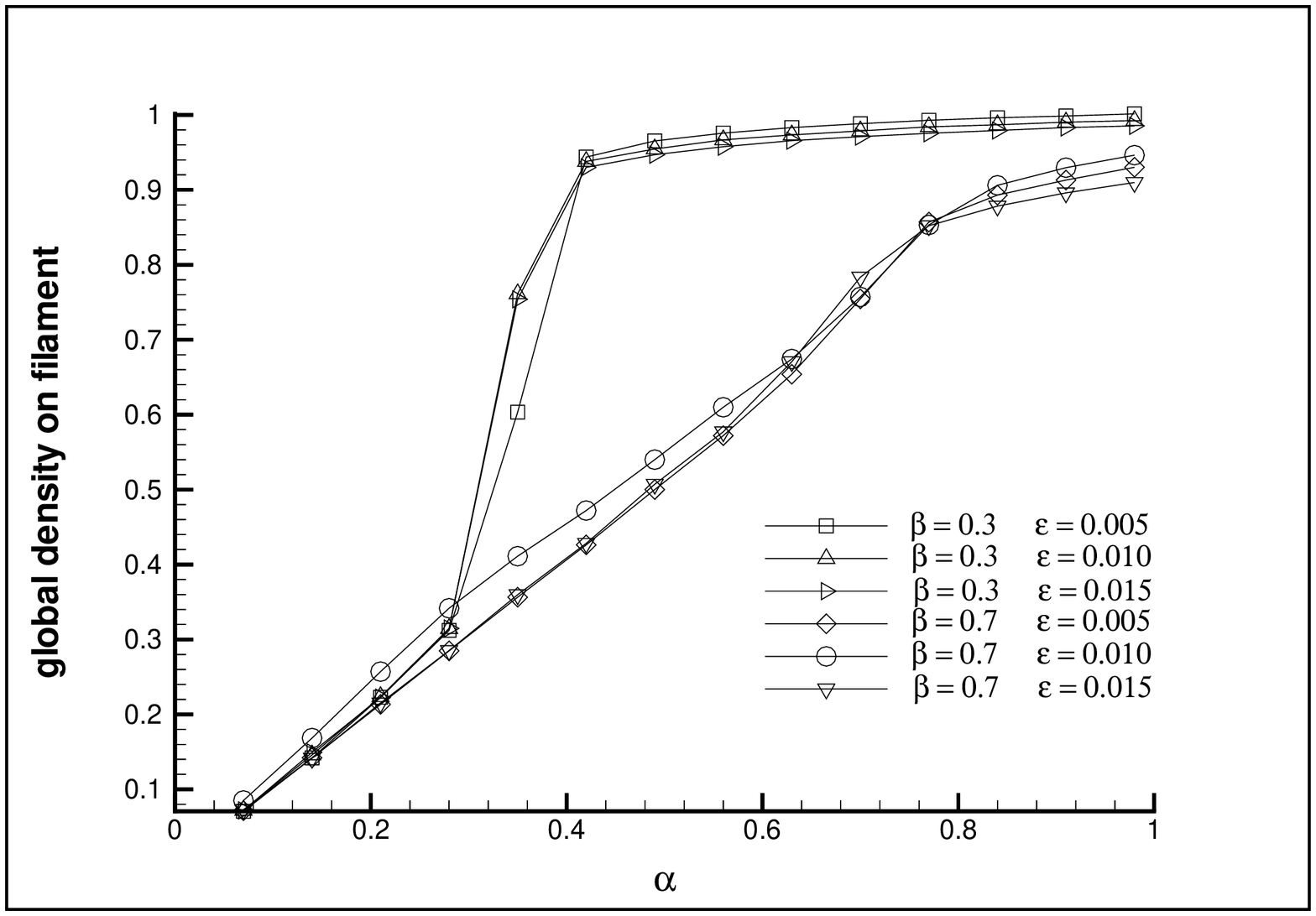}}
\end{figure}
\vspace{0.02 cm} {\small{Fig.~4: steady state $\rho_b$ in terms of
$\alpha$ for a system of size $Lx=100$. The values of $\beta$ and $\epsilon$ are specified in the figure.} }\\

Again, one observes phase transition behaviour from a low to a
high density phase for $\beta=0.3$ but a smooth crossover-like
behaviour for $\beta=0.7$. The results for $\beta=0.3$ i.e., a
first order low-to-high density transition is similar to that of
ASEP. However when $\beta=0.7$ one observes slight differences to
the behaviour of ASEP. Specifically, we see deviations from linear
dependence of $\rho_b$ vs $\alpha$ before reaching the saturation
region. We also obtained the above diagram for $Lx=200$ and $300$.
The results are very close to the case $Lx=100$. We next explore
the properties of the average position of the bound motors
$<x>_b$. The behaviour of $<x>_b$ in terms of $\alpha$ is
exhibited in the next graph.

\begin{figure}\label{Fig5}
\epsfxsize=8truecm \centerline{\epsfbox{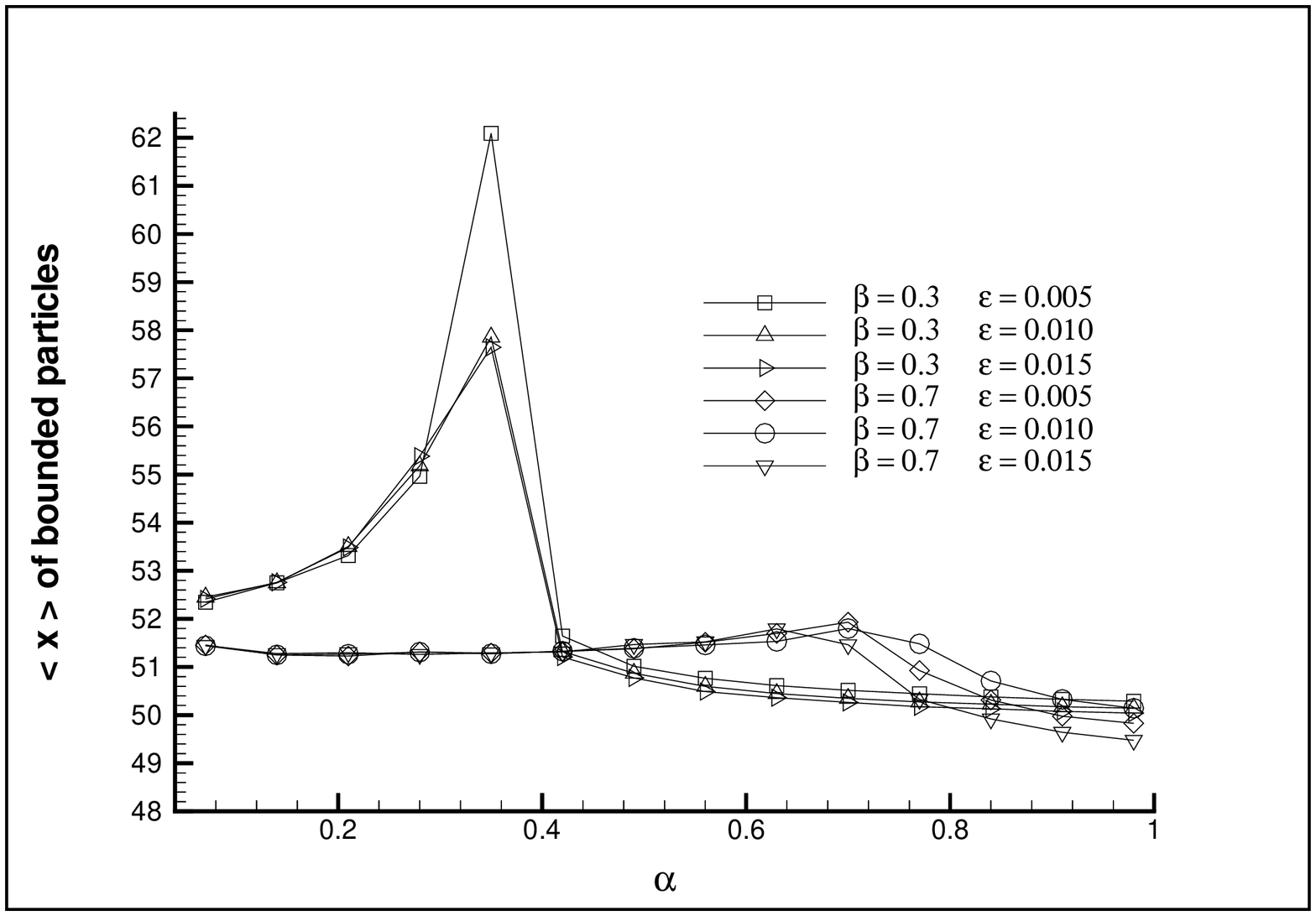}}
\end{figure}
\vspace{0.02 cm} {\small{Fig.~5: steady state $<x>_b$ in terms of
$\alpha$ for a system of size $Lx=100$. The value of $\beta$ and $\epsilon$ are specified in the figure.} }\\

It is interesting to note that $<x>_b$ reaches its peak at the
transition point but decreases afterwards. Before transition, the
system's load as well as $\rho_b$ are small and one does not
expect jam formation hence $<x>_b$ should be around half of the
filament length (here $50$). Near transition point, jams are
formed near the exit point and correspondingly $<x>_b$ increases
over $\frac{L_x}{2}=50$. The interesting point is that $<x>_b$
starts decreasing with further increase of $\alpha$. This could be
related to the bulk effects due to unbound particles. For
$\beta=0.7$ this effect is suppressed and the sharp peak is
smeared out. The appearance of a peak in the centre of mass of
bound particles signifies an important characteristics. The peak
tells us that around the value $\alpha \sim 0.35$ an inhomogeneous
density structure forms throughout the filament. This point will
become more clear by examining the filament density profile. To
get a deeper insight, it would be useful to consider density
profile on the filament. Figure (6) shows such profiles.

\begin{figure}\label{Fig6}
\epsfxsize=8truecm \centerline{\epsfbox{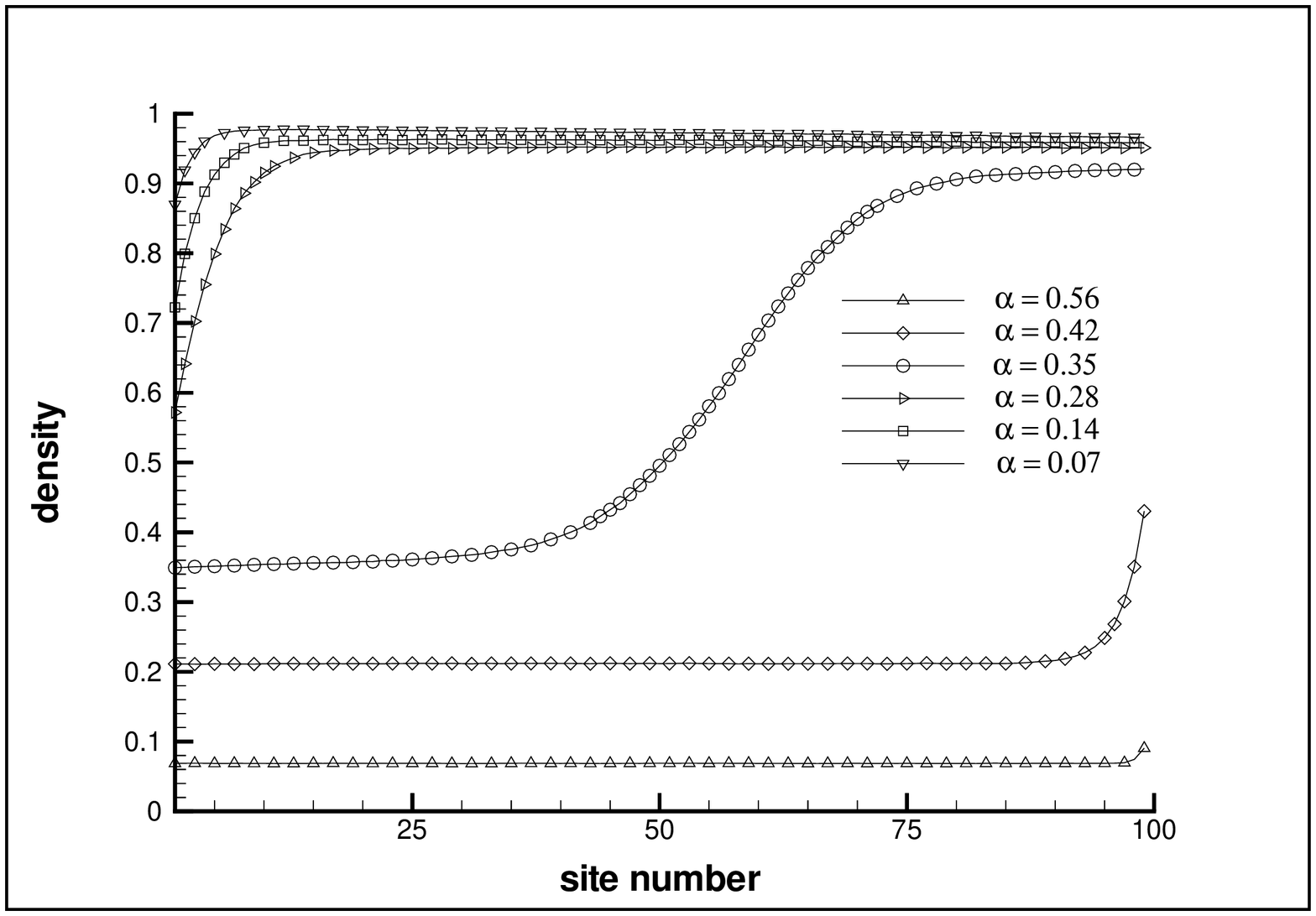}}
\end{figure}
\vspace{0.02 cm} {\small{Fig.~6: density profile on the filament
for various values of $\alpha$ for a system of size $Lx=100$. $\beta=0.3$ and $\epsilon=0.005$ .} }\\

For small values of $\epsilon$, we observe analogous behaviour to
ASEP. One has an exponential increase near the right boundary in
the low density phase. In marked contrast to ASEP, we see that for
$\alpha$ equal to $0.35$, there is a coexistence of low and high
density regions. In other words, there is a localized shock
structure. This has been earlier reported when the bulk was
modelled by constant attachment/detachment rates.
\cite{frey1,frey2,evans1}. Due to finite size of the system, the
domain wall is not so sharp. Upon increasing $\alpha$, the density
again becomes nearly constant in the bulk with an exponential
density decrease near the left boundary. To see the effect of
varying $\epsilon$, in the following figure, we have depicted the
density profile for the same values of $\alpha$ but with
$\epsilon=0.015$.

\begin{figure}\label{Fig7}
\epsfxsize=8truecm \centerline{\epsfbox{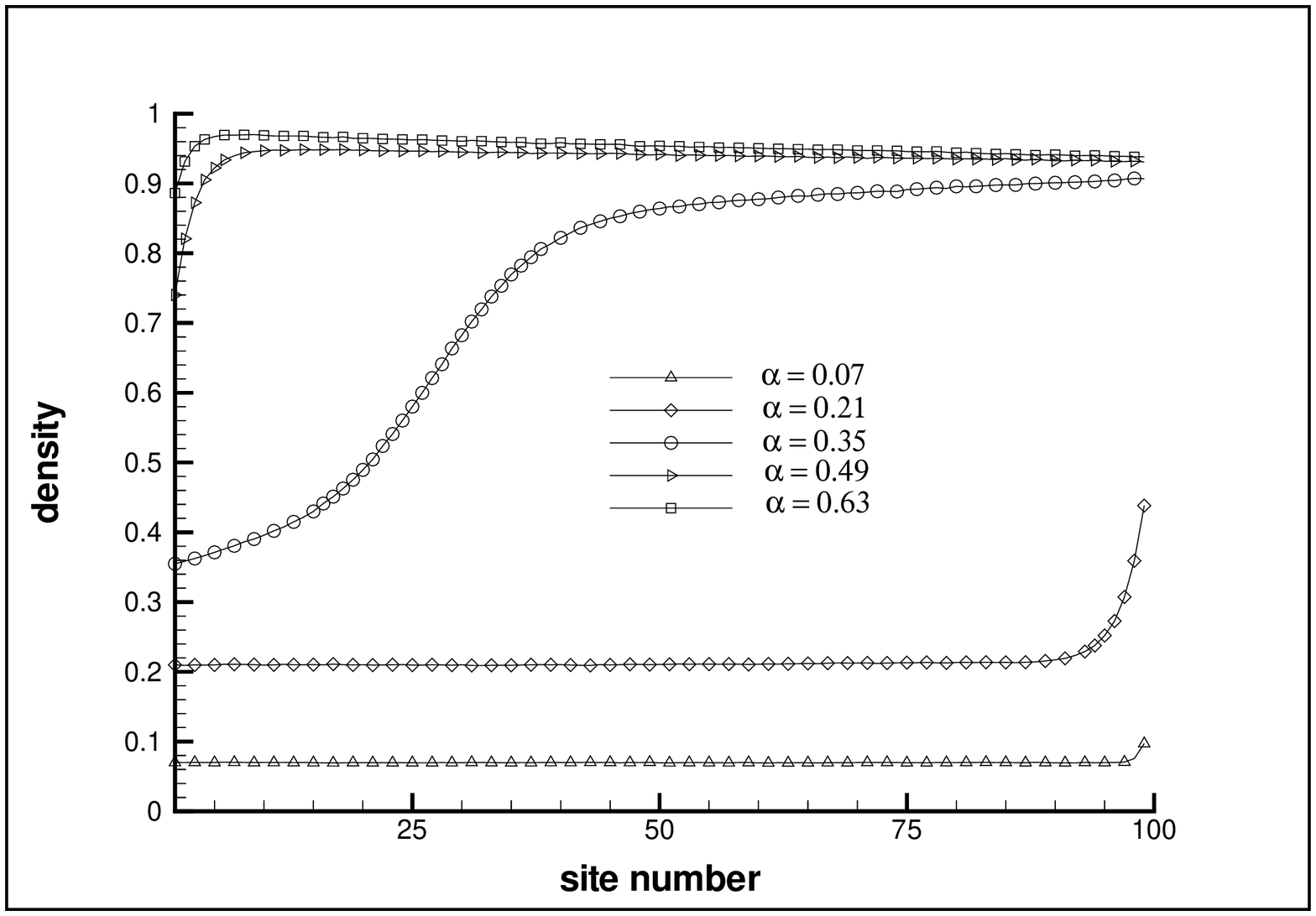}}
\end{figure}
\vspace{0.02 cm} {\small{Fig.~7: density profile on the filament
for various values of $\alpha$ for a system of size $Lx=100$, $\beta=0.3$ and $\epsilon=0.015$.} }\\

Analogous to the case $\epsilon=0.005$, we again have a localized
shock structure at the same $\alpha$ but this time the position of
the domain wall is shifted towards left. Furthermore, one can
recognize a linear profile, with a small slope, in the high
density regime. Although the slope is small but it is yet a
distinguishing aspect. This can be related to formation of
spontaneous jam or moving shocks and is a new feature of the
problem which is absent in ASEP. In fact, the interaction of one
dimensional filament with the unbound bulk particles leads to this
weakly linear density profile. Now let us investigate the model
properties when the exit rate $\beta$ is changed. The following
figure illustrates the behaviour of steady load vs $\beta$ for
$\alpha=0.3$ and $0.7$ each for three values of $\epsilon$.

\begin{figure}\label{Fig8}
\epsfxsize=8truecm \centerline{\epsfbox{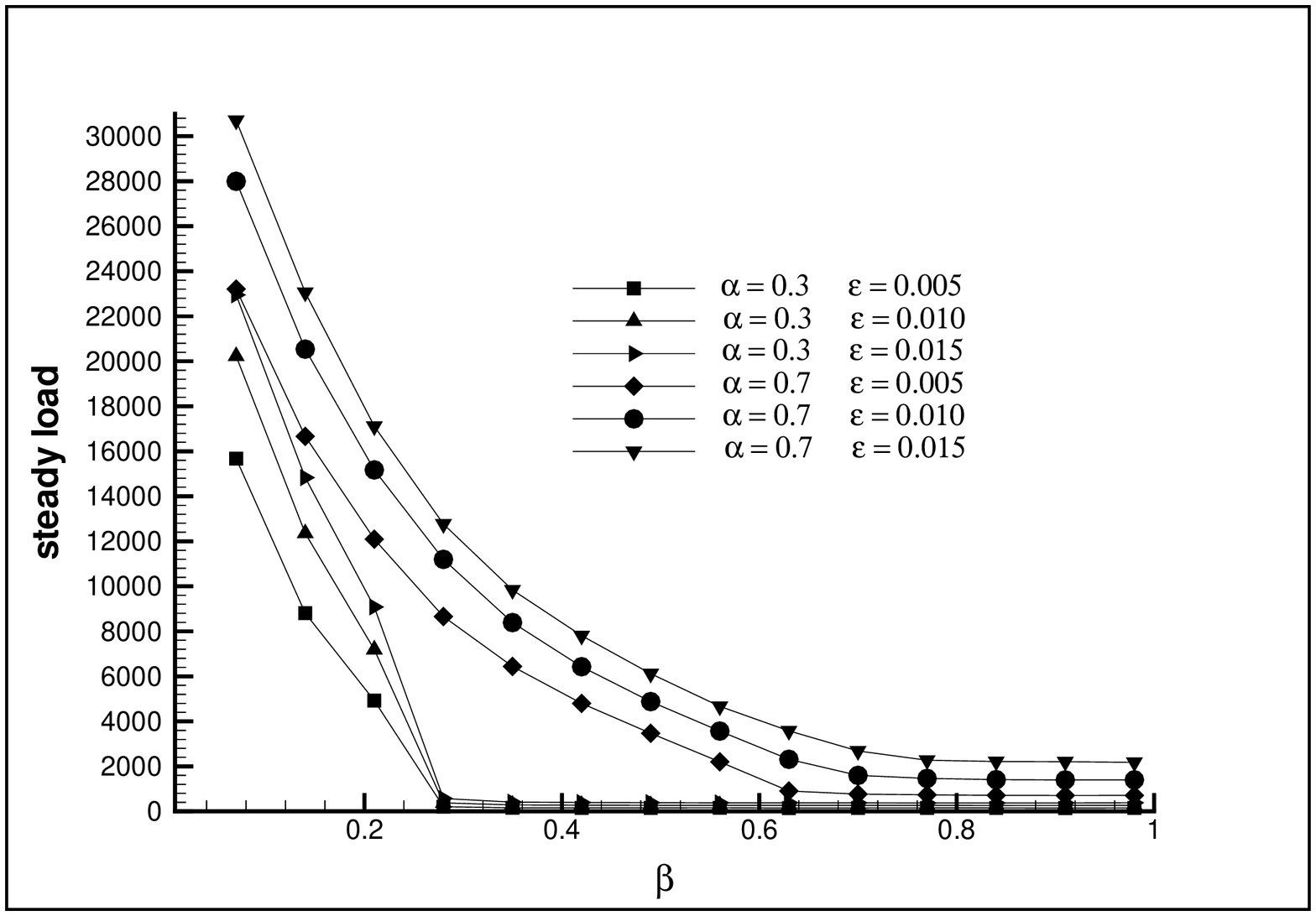}}
\end{figure}
\vspace{0.02 cm} {\small{Fig.~8: steady state load vs $\beta$ for
a
system of size $Lx=100$ for various values of $\alpha$ and $\epsilon$. } }\\

For $\alpha=0.3$ one observes a transition-like behaviour while
for larger $\alpha=0.7$ the abrupt behaviour is changed into a
smooth one. The effect of increasing $\epsilon$ is again smoothing
the transition behaviour. The passage time diagram is similar to
the load diagram and we do not show it. More interesting is the
steady current behaviour.

\begin{figure}\label{Fig9}
\epsfxsize=8truecm \centerline{\epsfbox{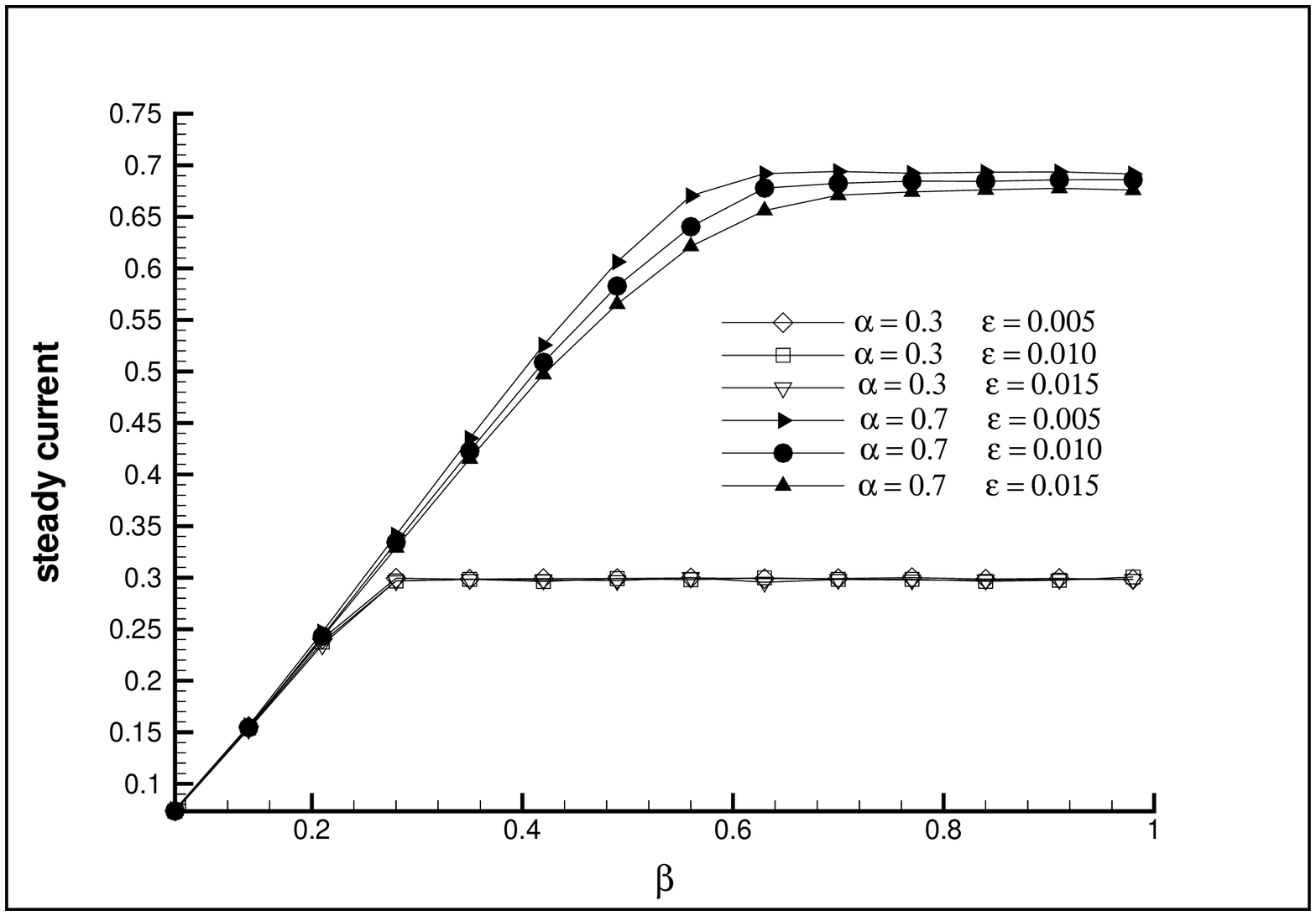}}
\end{figure}
\vspace{0.02 cm} {\small{Fig.~9: steady state current in terms of
$\beta$ for a system of size $Lx=100$. The value of $\alpha$ and
$\epsilon$ are the same as in the previous figure.} }\\

Analogous to ASEP, the current saturates at a $\alpha$-dependent
value. The effect of $\epsilon$ is notable only for high $\alpha$,
corresponding to high density phase, where its increase gives rise
to a current decline as expected. Besides this feature, for high
$\alpha$ we see that varying $\epsilon$ gives rise to smoothing of
the changes in the current. Next we look at the global density on
the filament. The following graph depicts this behaviour.

\begin{figure}\label{Fig10}
\epsfxsize=8truecm \centerline{\epsfbox{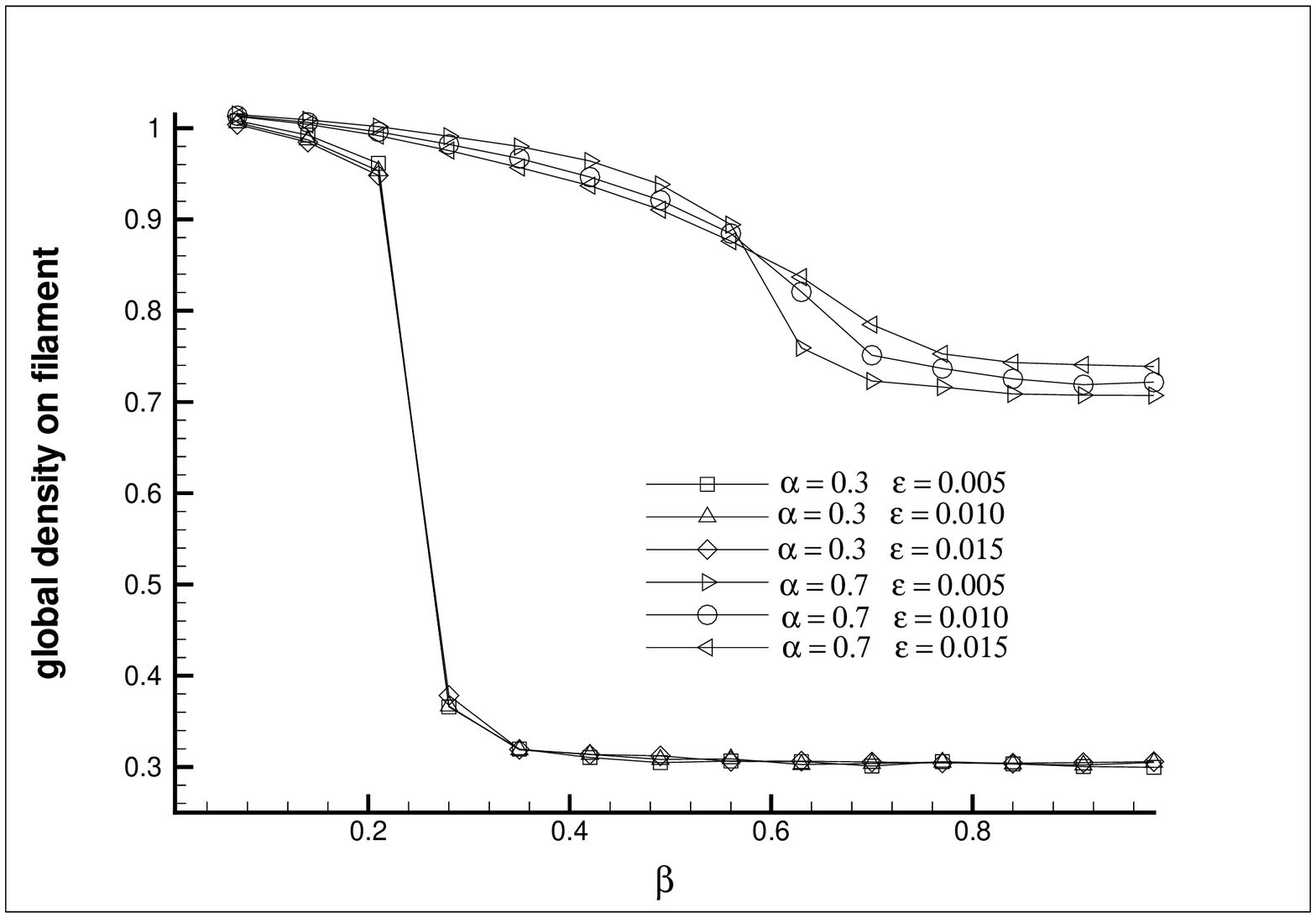}}
\end{figure}
\vspace{0.02 cm} {\small{Fig.~10: steady state $\rho_b$ in terms
of $\beta$ for a system of size $Lx=100$. The value of $\alpha$ and $\epsilon$ are specified in the figure.} }\\

Similar to ASEP, one observes a first order high-to-low density
transition for $\alpha=0.3$ and a smoothly decreasing behaviour
for $\alpha=0.7$ until it reaches a constant value. The transition
occurs at $\beta \sim 0.25$. We also examined the case $Lx=200$
and $300$. The results do not show significant changes to the case
$Lx=100$. This shows that the behaviour is not due to finite size
effects but is related to the bulk influence. We have also
evaluated the $<x>_b$ i.e., the average position of bound
particles vs $\beta$. Figure (11) shows this dependence.

\begin{figure}\label{Fig11}
\epsfxsize=8truecm \centerline{\epsfbox{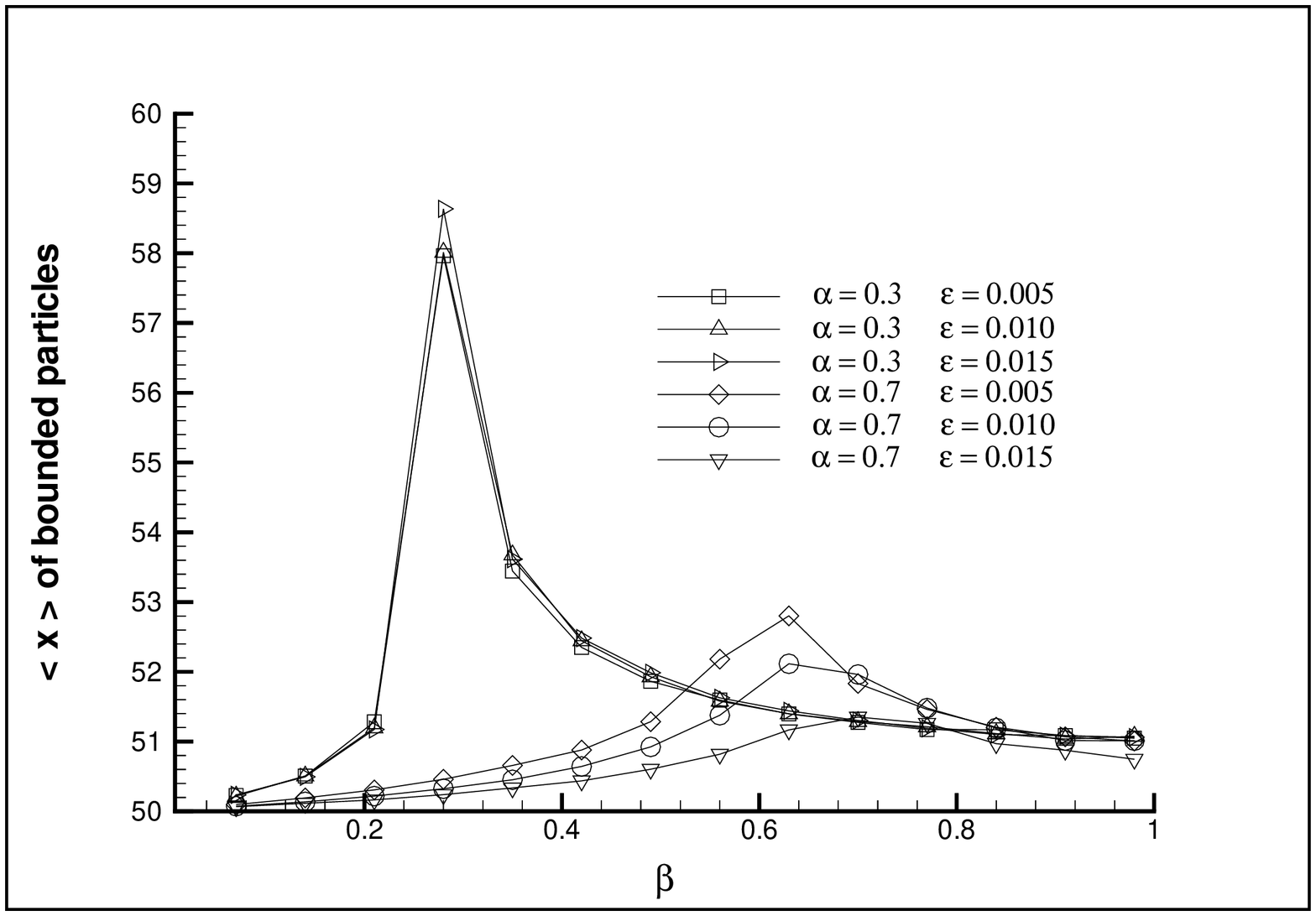}}
\end{figure}
\vspace{0.02 cm} {\small{Fig.~11: steady state $<x>_b$ in terms of
$\beta$ for a system of size $Lx=100$. The value of $\alpha$ and $\epsilon$ are specified in the figure.} }\\

Similar to figure (5), we see a sharp increase of $<x>_b$ to a
maximum value $\sim 0.6 Lx$ at the transition point. It decreases
to half of the filament for larger values of $\beta$. The position
of maximum crucially depends on $\alpha$ and $\epsilon$. For
larger values of $\alpha$, the maximum point shifts towards larger
$\beta$. Moreover, the maximum value of $<x>_b$ reduces. The
reason is that for larger $\alpha$, the system contains more
particles and can maintain high current. This leads to a more
homogenised density profile. Lastly, we depict the density
profiles for some values of $\beta$ with $\epsilon=0.01$ and
$0.015$ in the following figures. In comparison to ASEP, we see
entirely different profiles. Generally, the profile is increased
and then starts declining in a linear manner.

\begin{figure}\label{Fig12}
\epsfxsize=8truecm \centerline{\epsfbox{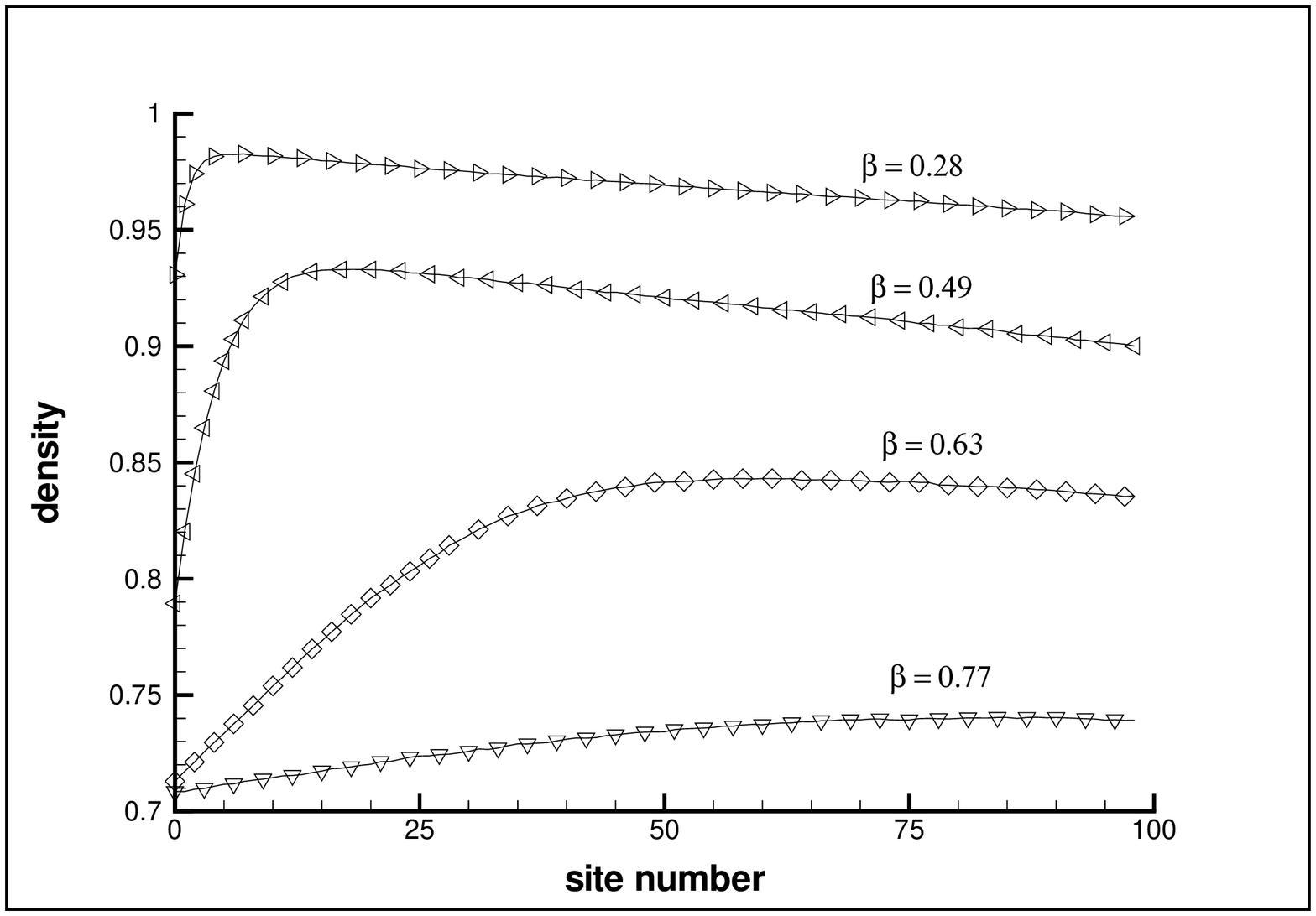}}
\end{figure}
\vspace{0.02 cm} {\small{Figs.~12: density profile on the filament
for various values of $\beta$ for a system of size $Lx=100$ and
$\alpha=0.7$. $\epsilon=0.01$.} }\\

\begin{figure}\label{Fig13}
\epsfxsize=8truecm \centerline{\epsfbox{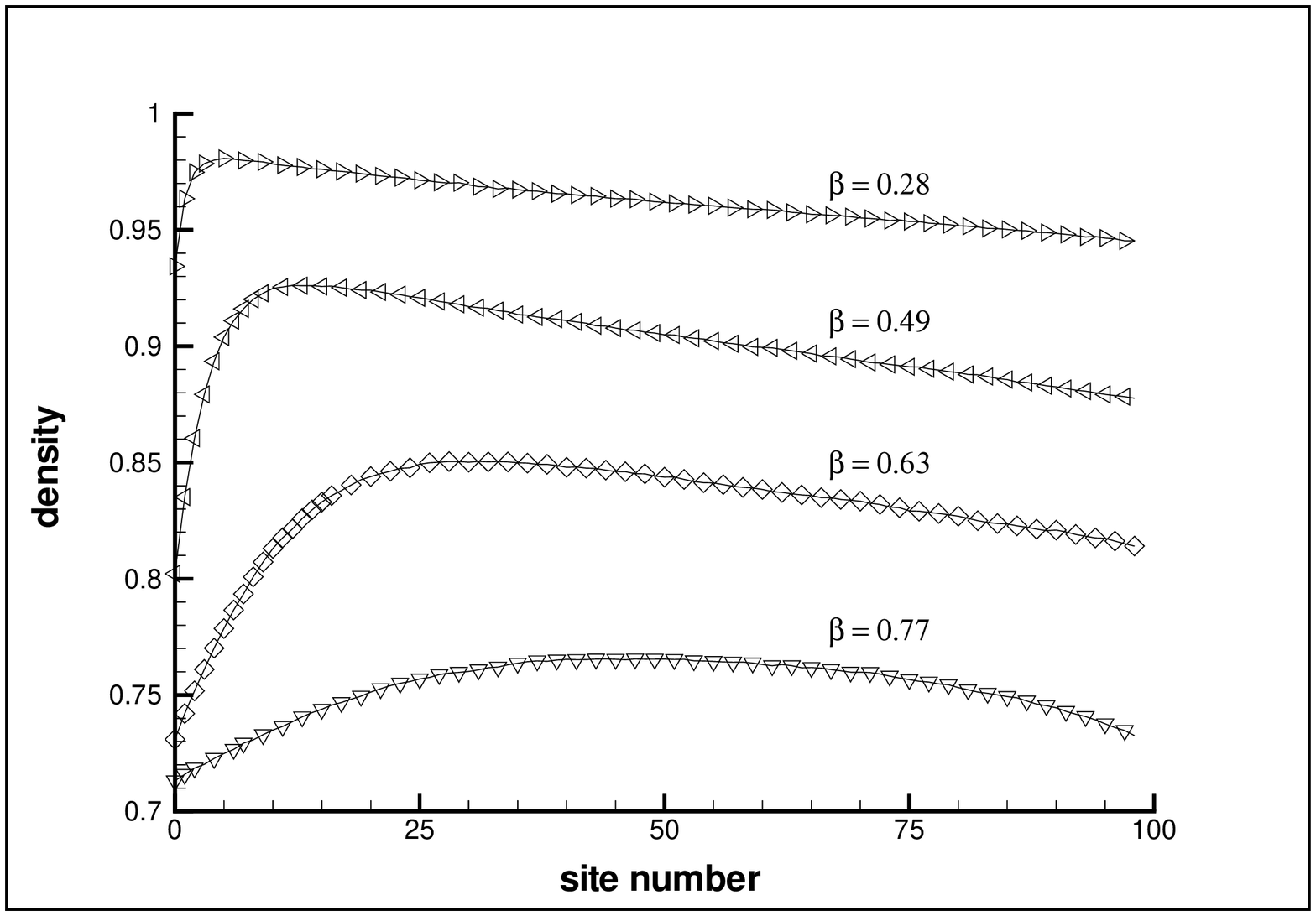}}
\end{figure}
\vspace{0.02 cm} {\small{Figs.~13: density profile on the filament
for various values of $\beta$ for a system of size $Lx=100$ and
$\alpha=0.7$. $\epsilon=0.015$.} }\\

The effect of increasing $\beta$ is two fold. First, smoothing the
changes between increasing and decreasing regions of the density
profile and second, extending the region of increasing profile. In
comparison to the top figure where $\epsilon=0.01$, for a larger
value of $\epsilon=0.015$ we see that the size of region where the
profile is increasing, decreases. Increasing $\epsilon$ is more
effective when the exit rate $\beta$ is large. For instance, when
$\beta=0.63$, the increasing part of the density profile decreases
in comparison to the case $\epsilon=0.005$.

To see the behaviour of the density profile for larger systems
size, we also obtained the profile for the same parameters as in
the above figures but this time for $Lx=300$. The following figure
depicts the behaviour:

\begin{figure}\label{Fig14}
\epsfxsize=8truecm \centerline{\epsfbox{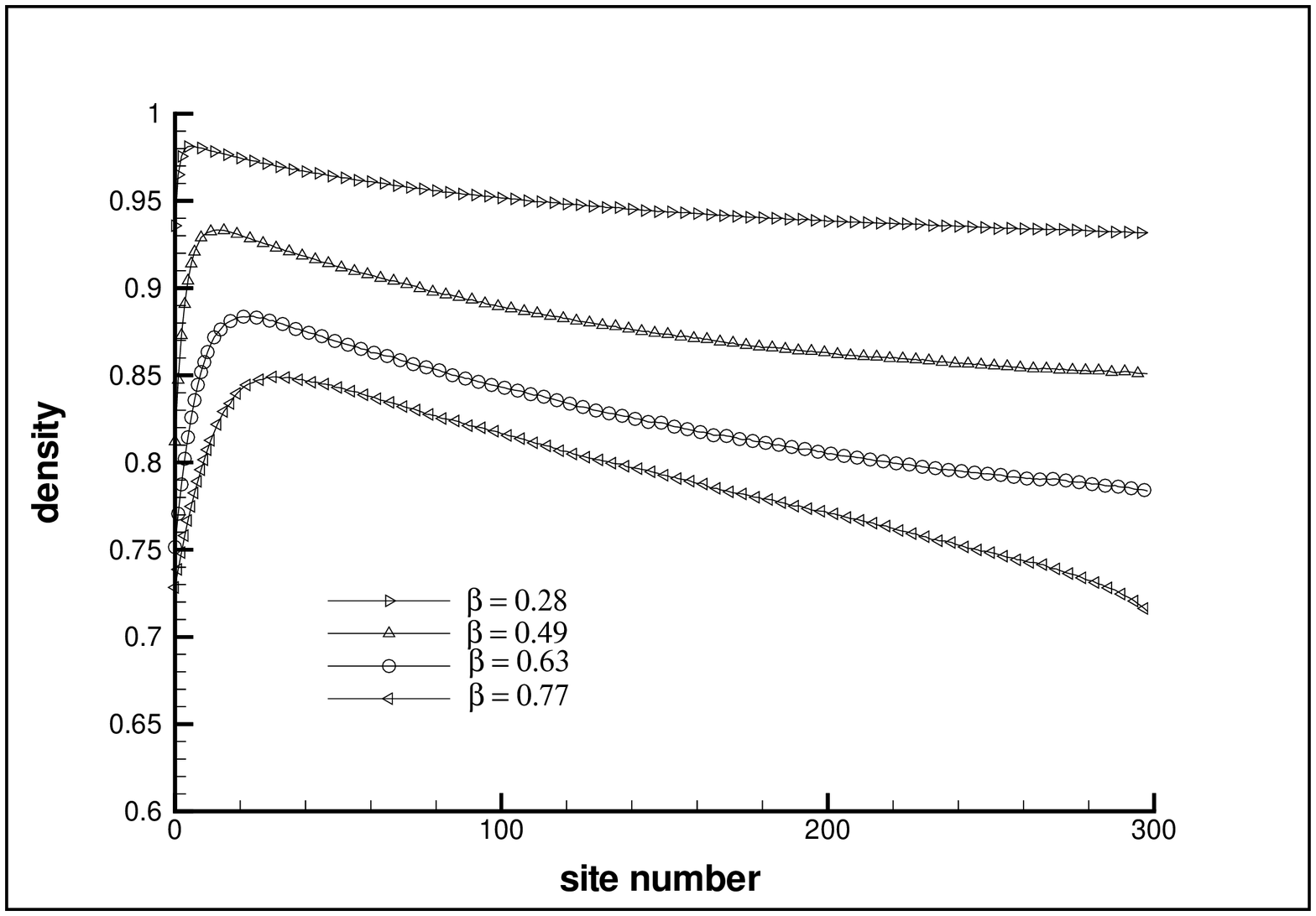}}
\end{figure}
\vspace{0.02 cm} {\small{Fig.~14: density profile on the filament
for various values of $\beta$ for a system of size $Lx=300$,
$\alpha=0.7$ and $\epsilon=0.015$.} }\\

One observes that for large $\beta$ a linear profile forms in the
bulk. The effect of the left boundary on the profile is the
formation of an increasing density layer. The location of the
maximum point moves away the left boundary for higher $\beta$. One
also observes a concave density profile structure for sufficiently
small $\beta$. This aspect was absent in $Lx=100$ where not only
the concavity did not exist but also for $\beta=0.77$ there
existed a concave profile. In order to have a deeper insight to
the problem, we looked in more details into the effect of
$\epsilon$ on the density profile. In the following figures, we
show the density profile for some values of $\epsilon$ . $\alpha$
and $\beta$ are fixed. We first consider the case
$\alpha=\beta=0.7$.

\begin{figure}\label{Fig15}
\epsfxsize=8truecm \centerline{\epsfbox{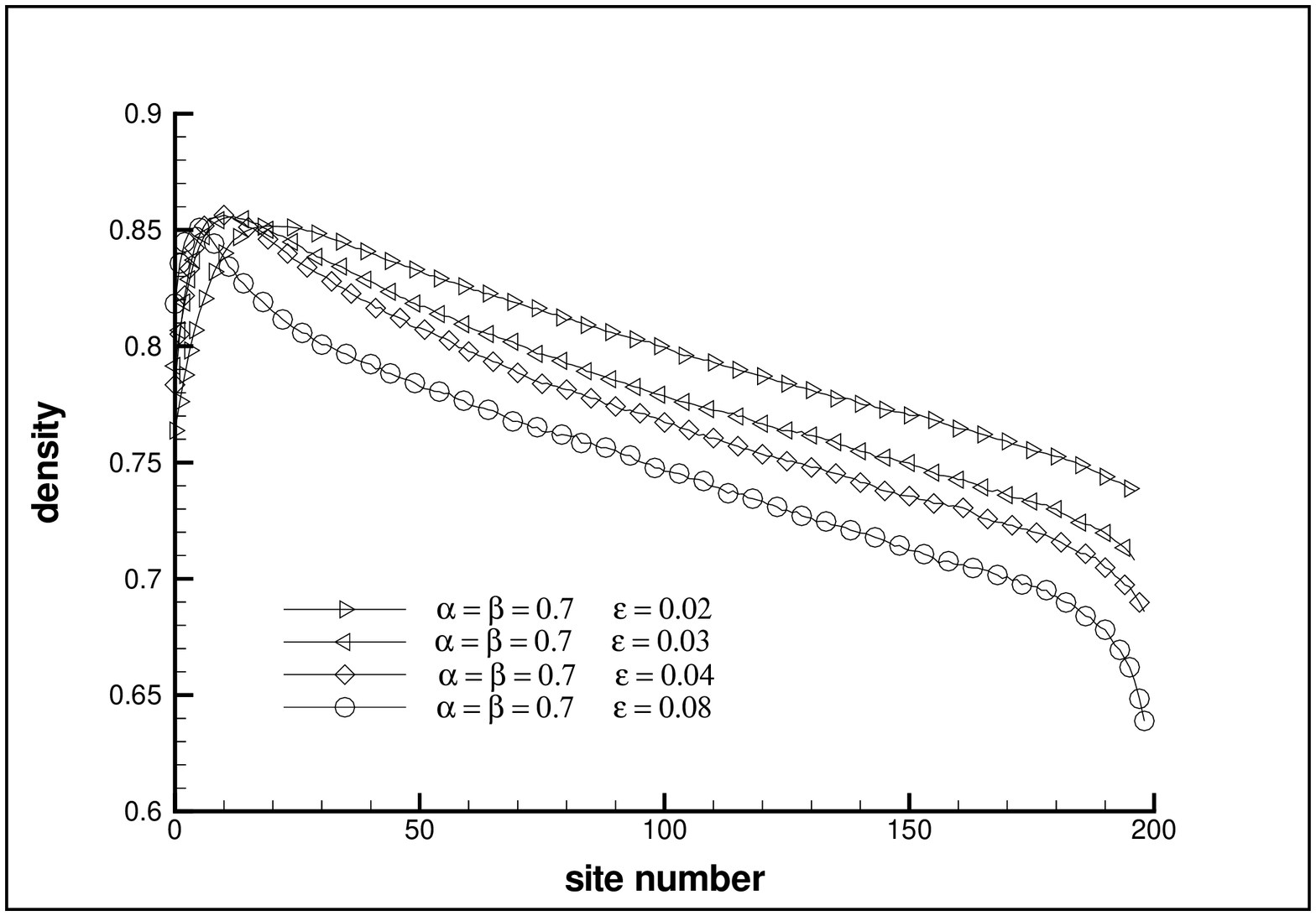}}
\end{figure}
\vspace{0.02 cm} {\small{Fig.~15: density profile on the filament
for various values of $\epsilon$ for a system of size $Lx=200$,
$Ly=Lz=25$, $\alpha=0.7$ and $\beta=0.7$.} }\\

Except, a short region where the profile is increasing, we again
observe a linearly decreasing profile. The overall effect of
increasing $\epsilon$ is to reduce the density value throughout
the chain. The reason for such decrease in the local densities
could possibly be explained on account that in the present case
$\alpha=\beta$ the system can maintain a large current therefore
increasing $\epsilon$ would effectively reduces the number of
bound particles and therefore the local density on the filament
decreases. The next figure exhibits the density profile for the
case $\alpha=\beta=0.3$ .

\begin{figure}\label{Fig16}
\epsfxsize=8truecm \centerline{\epsfbox{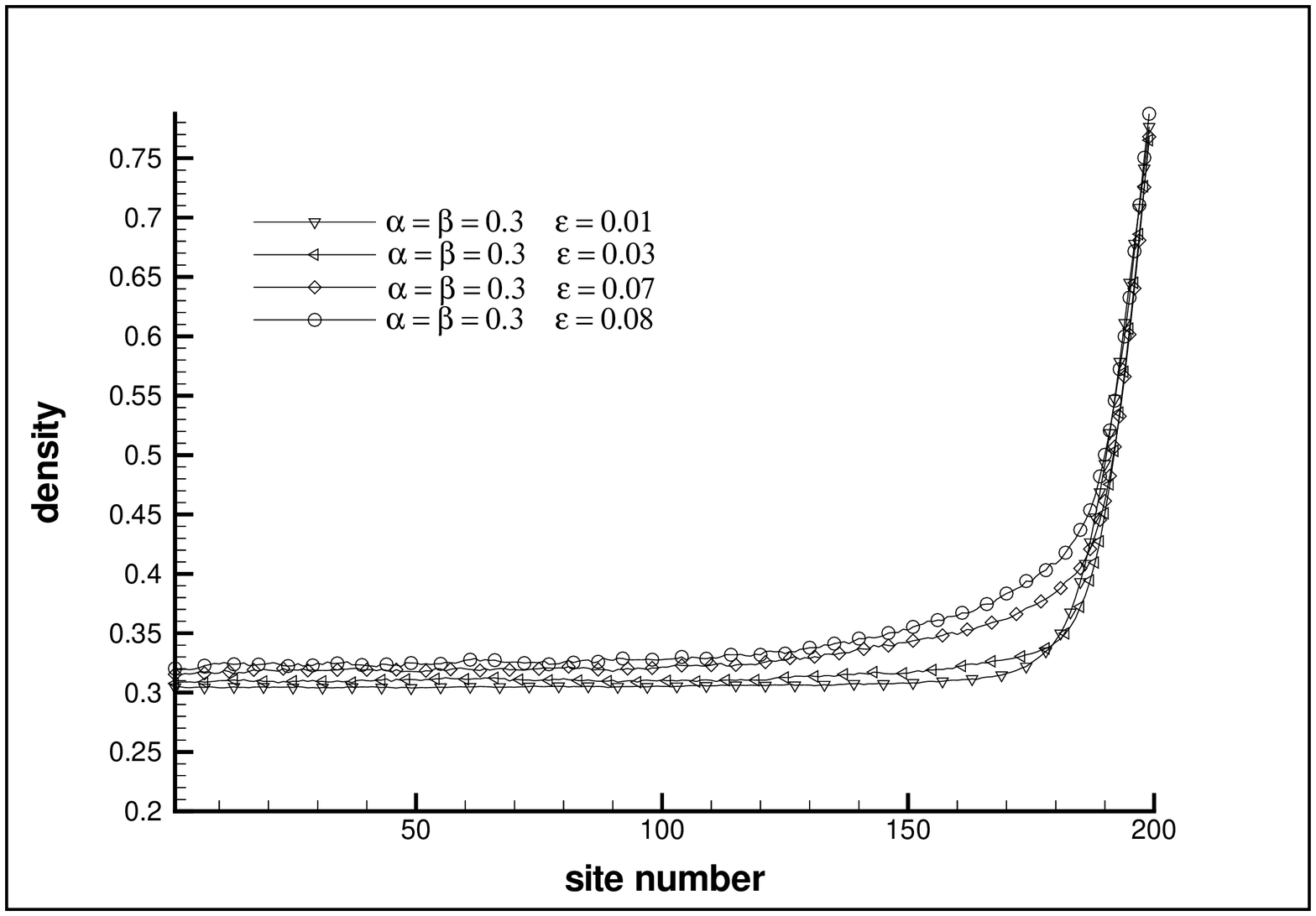}}
\end{figure}
\vspace{0.02 cm} {\small{Fig.~16: density profile on the filament
for various values of $\epsilon$ for a system of size $Lx=200$,
$Ly=Lz=25$, $\alpha=0.3$ and $\beta=0.3$.} }\\

Here we observe a constant profile in the bulk accommodated with
sharp increasing layer at the right boundary. This is also in
contrast to ASEP where once expects a linear profile along the
$\alpha=\beta \leq \frac{1}{2}$ line. We conclude that the
the detachment rate $\epsilon$ can give rise to
significantly different features which are absent in ASEP. This
conclusion have been earlier obtained for similar models without
taking into account the dynamic of unbound particles
\cite{frey1,frey2,evans1,juhasz,klump6}. Our final graph depicts
the density profile for some values of larger $\epsilon$. $\alpha$
and $\beta$ are fixed but unequal.

\begin{figure}\label{Fig17}
\epsfxsize=8truecm \centerline{\epsfbox{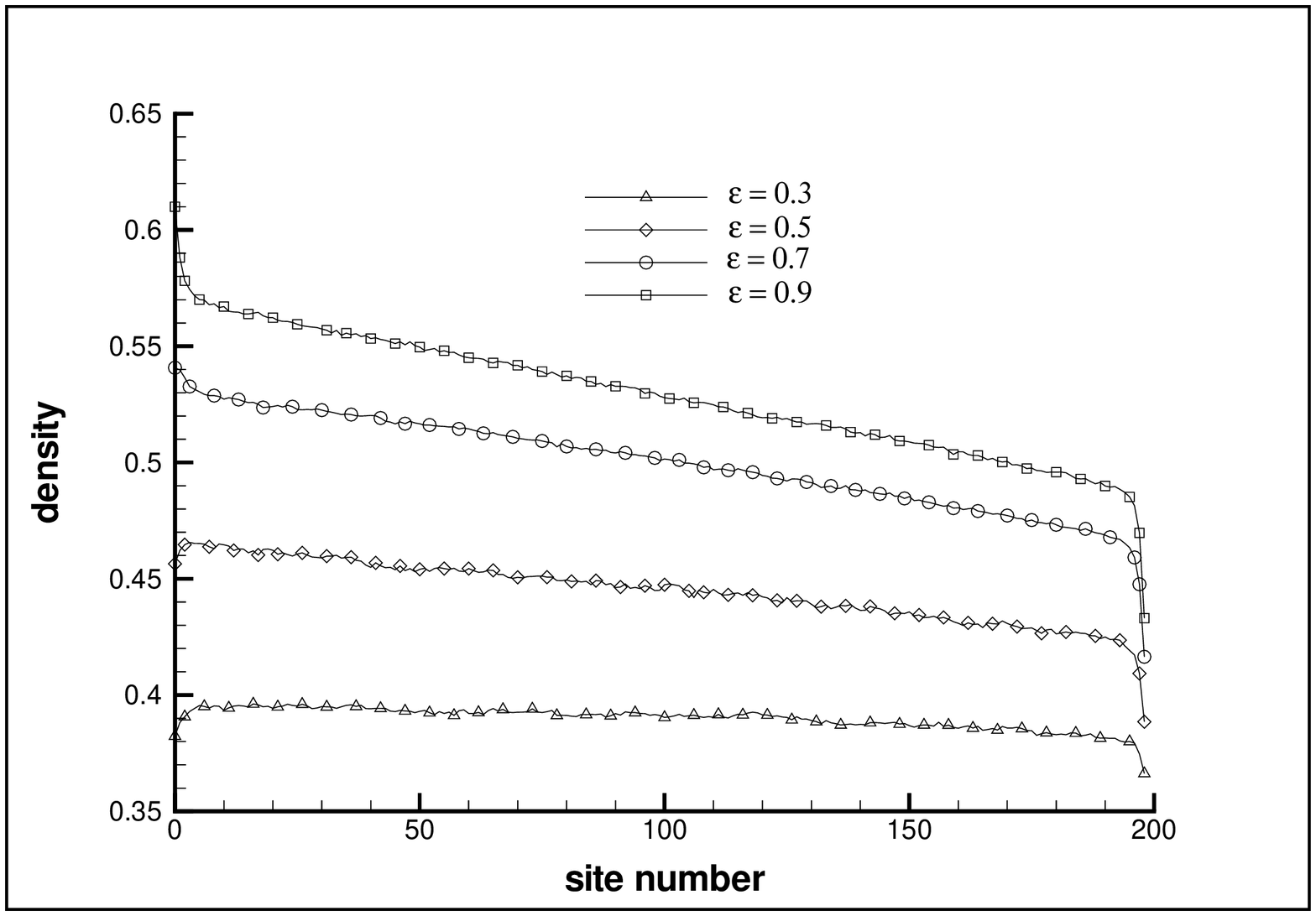}}
\end{figure}
\vspace{0.02 cm} {\small{Fig.~17: density profile for various
values of $\epsilon$ for a system of size $Lx=200$,
$Ly=Lz=25$, $\alpha=0.3$ and $\beta=0.7$.} }\\

The above graph exhibits the profile in a low congestion state
corresponding to $\alpha=0.3$ and $\beta=0.7$. As observed, the
effect of $\epsilon$ is more notable for higher values. For small
$\epsilon$, the effect is an upward shift in the density profile.
Larger $\epsilon$ not only shifts the density profile towards a
higher value but also creates a slope in it. Finally, to see the
effect of lateral dimensions of the compartment i.e., $Ly$ and
$Lz$, we performed extensive Monte Carlo simulations for
$Ly=Lz=12$ up to $70$. Our results showed no significant changes
for the case where $\epsilon$ is small.

\section{Summary and Concluding Remarks}

Let us now summarize what has been done in this paper. We have
developed a discrete time cellular automata model for the
description of motor protein traffic through a filament-like
track. Motors execute totally asymmetric random walk when bound
onto the track. They randomly detach from the filament and perform
unbiased random walk inside a closed compartment with reflecting
boundary conditions. A similar problem has been earlier considered
in \cite{klump3}. Our boundary condition and updating schemes is
different to that in \cite{klump3}. Our main interest has been the
interplay of the bulk and the boundary on the stationary
properties of the system. In particular we have studied the effect
of varying the detachment rate $\epsilon$ on the transport
characteristics of the model. Although the model behaviours
resemble that of ASEP in some ranges of parameters, especially
when the unbinding rate is sufficiently low, Our investigation has
illustrated that for considerable high values of the unbinding
rate, one observes remarkable distinct features with respect to
ASEP. Specifically, the density profile shows linear and localized
behaviour in some ranges of parameters. This implies that the
density shock structure of the models should be entirely different
than ASEP. This aspect can be of relevance to biological
applications. Besides biological motivation, our study shed more
lights onto the problem of bulk-boundary interplay in the
transport characteristics of low dimensional non-equilibrium
systems which recently has attracted the attention of researchers
in the field. Analogous to the cases where the
detachment/attachment rates are constant
\cite{frey1,frey2,evans1}, we see the coexistence of high and low
density regions which is termed {\it shock localization}.
Moreover, we also observe other types of density structures such
as linear density profiles with distinctive boundary behaviours.
This shows that incorporating dynamics for the bulk particles, can
drastically affect the phase structure of the system. It would be
illustrative to find out the phase diagram structure of the model.
This is our next step. Finally we recall that our model does not
incorporate the internal degrees of freedom for motors. In
practice, these motors resemble a two-headed creatures. Taking
this fact into a quantitative description \cite{kolomeisky2}
together with chemical state of the motors are two crucial steps
which should be considered in further progresses \cite{schadprl}.
Work along this line is in preparation and will be reported
elsewhere.

\section{acknowledgement}

We highly wish to acknowledge the {\it Institute of Advanced
Studies in Basic Sciences} (IASBS) for proving us with the
computational facilities where the final stages of this work were
carried out.

\bibliographystyle{unsrt}

\end{multicols}

\end{document}